\documentclass{emulateapj}

\usepackage{amsmath}
\usepackage{amssymb}
\usepackage{graphicx}
\usepackage{color}
\usepackage{natbib}
\usepackage{epsfig}
\def\gtaprx {\lower .1ex\hbox{\rlap{\raise .6ex\hbox{\hskip .3ex
	{\ifmmode{\scriptscriptstyle >}\else
		{$\scriptscriptstyle >$}\fi}}}
	\kern -.4ex{\ifmmode{\scriptscriptstyle \sim}\else
		{$\scriptscriptstyle\sim$}\fi}}}
\def\ltaprx {\lower .1ex\hbox{\rlap{\raise .6ex\hbox{\hskip .3ex
	{\ifmmode{\scriptscriptstyle <}\else
		{$\scriptscriptstyle <$}\fi}}}
	\kern -.4ex{\ifmmode{\scriptscriptstyle \sim}\else
		{$\scriptscriptstyle\sim$}\fi}}}

\newcommand{\cutt}[1]{\textcolor{blue}{}}

\newcommand{\Ms}{{\ensuremath{\mathrm{M}_{\odot} }}}

\newcommand{\Ni}{{\ensuremath{^{56}\mathrm{Ni}}}}

\begin{document}

\title{Seeing the First Supernovae at the Edge of the Universe with JWST}

\author{Daniel J. Whalen\altaffilmark{1}, Chris L. Fryer\altaffilmark{2}, Daniel E. Holz\altaffilmark{3}, 
Alexander Heger\altaffilmark{4}, S. E. Woosley\altaffilmark{5}, Massimo Stiavelli\altaffilmark{6}, 
Wesley Even\altaffilmark{7} and Lucille H. Frey\altaffilmark{7,8}}

\altaffiltext{1}{McWilliams Fellow, Department of Physics, Carnegie Mellon 
University, Pittsburgh, PA 15213}

\altaffiltext{2}{CCS-2, Los Alamos National Laboratory, Los Alamos, NM 87545}

\altaffiltext{3}{Enrico Fermi Institute, Department of Physics, and Kavli Institute for 
Cosmological Physics, University of Chicago, Chicago, IL 60637, US}

\altaffiltext{4}{Monash Centre for Astrophysics, Monash University, Victoria, 3800, Australia}

\altaffiltext{5}{Department of Astronomy and Astrophysics, UCSC, Santa Cruz, CA  
95064}

\altaffiltext{6}{Space Telescope Science Institute, 3700 San Martin Drive, Baltimore, MD 21218}

\altaffiltext{7}{XTD-6, Los Alamos National Laboratory, Los Alamos, NM 87545}

\altaffiltext{8}{Department of Computer Science, University of New Mexico, Albuquerque, NM  87131}

\begin{abstract}

The first stars ended the cosmic Dark Ages and created the first heavy elements necessary for 
the formation of planets and life. The properties of these stars remain uncertain, and it may be 
decades before individual Pop III stars are directly observed.  Their masses, however, can be 
inferred from their supernova explosions, which may soon be found in both deep-field surveys 
by the \textit{James Webb Space Telescope} (\textit{JWST}) and in all-sky surveys by the 
\textit{Wide Field Infrared Survey Telescope} (\textit{WFIRST}).  We have performed radiation 
hydrodynamical simulations of the near infrared signals of Pop III pair-instability supernovae in 
realistic circumstellar environments with Lyman absorption by the neutral intergalactic medium.  
We find that \textit{JWST} and \textit{WFIRST} will detect these explosions out to $z \sim 30$ 
and 20, respectively, unveiling the first generation of stars in the universe.

\end{abstract}


\maketitle

\section{Introduction}

The cosmic Dark Ages ended with the formation of the first stars in 10$^5$--10$^6$ \Ms\ 
cosmological halos at $z \sim$ 20--30.  Besides ionizing the universe with intense UV flux and 
forming the first heavy elements, Pop III stars also determined the luminosities and spectra of 
primeval galaxies and may have been the origin of the supermassive black holes (SMBHs) at 
the centers of most massive galaxies today.  Unfortunately, because they lie at the edge of 
the observable universe, individual Pop III stars will likely remain beyond the reach of direct 
observation for decades to come despite their enormous luminosities \citep{s02} and new 
near infrared (NIR) observatories such as \textit{JWST} \citep{jwst06} and the \textit{Thirty
Meter Telescope} ({TMT}). Past numerical simulations have suggested that Pop III stars form 
in isolation in halos, with masses of 30--300 \Ms\ \citep{nu01,bcl02,abn02,on07}, but more 
recent calculations point to the possibility of binaries \citep{turk09}, and perhaps small swarms 
of less massive 20--40 \Ms\ stars \citep{stacy10,clark11,get11,hos11,get12}.  However, in spite 
of their increasing sophistication numerical simulations cannot yet constrain the properties of 
the first stars, as they do not realistically bridge the gap between the formation and 
fragmentation of protostellar disks and their destruction by nascent stars up to a Myr later.

It may soon be possible to deduce the properties of Pop III stars from their supernova (SN) 
explosions.  Primordial stars from 10--40 \Ms\ are thought to die in core-collapse SNe while
140--260 \Ms\ stars explode as pair-instability (PI) SNe, extremely energetic thermonuclear 
explosions with up to 100 times the energy of Type Ia or II SNe \citep{hw02,wet08a,jet09b,
hw10,jw11}.  Pop III stars above 100 \Ms\ encounter the pair instability after central carbon 
burning, when thermal energy creates electron-positron pairs rather than providing pressure 
support against collapse \citep{brk67}. Their cores subsequently contract, triggering explosive 
thermonuclear burning of O and Si.  Above 140 \Ms\ the energy released completely unbinds 
the star, with no compact remnant.  At 260 \Ms\ the core becomes so hot that alpha particles 
are photodisintegrated into free nucleons, consuming as much energy per unit mass as was 
released by all preceding burning, and the star collapses instead of exploding.  PI SNe such 
as SN 2007bi have now been discovered in the local universe \citep{gy09}, and the recent 
discovery that rotating Pop III stars can die as PI SNe at masses as low as 65 \Ms\ may 
increase the number of such explosions at high redshifts by a factor of a few \citep{cw12}.

Pop III PI SNe could be ideal probes of the primeval universe because they are 100,000 times 
brighter than their progenitors and the early galaxies in which they reside, and might be found 
by \textit{JWST} or in all-sky surveys by \textit{WFIRST} or the \textit{Wide-field Imaging 
Surveyor for High-Redshift} (\textit{WISH}).  However, to determine if these events can be seen 
from Earth, fully radiation-hydrodynamical simulations of PI SNe must be performed in realistic 
circumstellar envelopes, whose opacity and interaction with the shock crucially shape the
temperatures and spectra in the source frame, and thus the flux that is eventually redshifted 
into the NIR.  Lyman absorption by neutral hydrogen prior to the era of reionization, which can 
absorb or scatter light from these ancient supernovae out of our line of sight, must also be 
taken into account for explosions at the earliest epochs.  Previous simulations of PI SNe have 
not included these effects \citep{sc05,kasen11}, and estimates of the redshift at which they can 
be detected have either been confined to $z <$ 15 (the era of primitive galaxies rather than the 
epoch of first light) or have been attempted at $z \sim$ 30 with only very approximate methods 
\citep{pan12a,pan12b,hum12}. 

We have performed numerical simulations of the NIR light curves of Pop III PI SNe.  In $\S \, 2$ 
we describe our numerical method and in $\S \, 3$ we present source frame light curves and 
spectra for 5 PI explosions.  In $\S \, 4$ NIR light curves for these events are shown for several 
redshifts, and we discuss their detection limits in redshift.

\begin{deluxetable}{ccccc}  
\tabletypesize{\scriptsize}  
\tablecaption{Pop III PI SN properties (masses are in \Ms\label{tab:table1})}
\tablehead{
\colhead{M$_{\star}$} & \colhead{$M_{He}$}& \colhead{$R$ ($10^{14}$ cm)}& \colhead{$E$ 
(10$^{51}$ erg)} & \colhead{$M_{\Ni}$}}
\startdata 
150  &  72      &  1.62    &   9.0    &  0.07  \\
175  &  84.4   &  1.74    &   21.3  &  0.70  \\
200  &  96.7   &  1.84    &   33     &  5.09  \\
225  &  103.5 &  3.33    &   46.7  &  16.5  \\
250  &  124    &  2.25    &   69.2  &  37.9      
\enddata 
\end{deluxetable}  

\section{Numerical Method}

Our calculations are done in four stages.  First, we model the evolution and initial explosion of 
150, 175, 200, 225, and 250 \Ms\ Pop III stars in the Kepler code \citep{Weaver1978,
Woosley2002}.  Second, the shock is propagated through the interior of the star, its surface, 
and then out into the surrounding medium with the radiation hydrodynamics code RAGE \citep{
rage}.  We model radiation flows with grey flux-limited diffusion and evolve matter and radiation 
temperatures separately because they are not always in equilibrium.  This physics enables us to 
better capture shock breakout \citep[e.g.][]{mm99,ns10,Piro10,Katz12}, as discussed in greater 
detail in \citep{fet12}.  Energy deposition in the gas due to the radioactive decay of \Ni, which 
powers the light curve at later times, is also included.

We surround the star with a power-law wind profile with a total mass of 0.1 \Ms, in keeping with 
the general consensus that massive zero-metallicity stars do not exhibit much mass loss due to 
their pristine atmospheres \citep{Vink01}.  The wind profile is smoothly joined to the H II region 
density profile of a massive Pop III star calculated with the ZEUS-MP code \citep{wan04,wn06}. 
ZEUS-MP also captures the ionization state of the wind due to UV radiation from the star, and 
therefore its opacity to radiation from the SN.  Because all five stars become red hypergiants 
and cease to emit ionizing radiation in the final few hundred kyr of their lives, the wind completely 
recombines by the time they explode while the low-density H II region beyond it remains ionized.

Third, we construct spectra from the RAGE blast profiles with the SPECTRUM code \citep{
fet12} using atomic opacities from the Los Alamos OPLIB database \citep{oplib}.  SPECTRUM 
calculates continuum and line luminosities, absorption lines imprinted on the flux by the ejecta 
and envelope, and Doppler shifts and time dilation due to relativistic expansion.  Finally, we 
convolve our spectra with Lyman absorption by neutral clouds \citep{madau95,su11}, 
cosmological redshifting, and filter response functions to calculate NIR light curves.  We 
summarize the properties of our explosions in Table 1.

\section{Source Frame Light Curves / Spectra}

\begin{figure}
\plotone{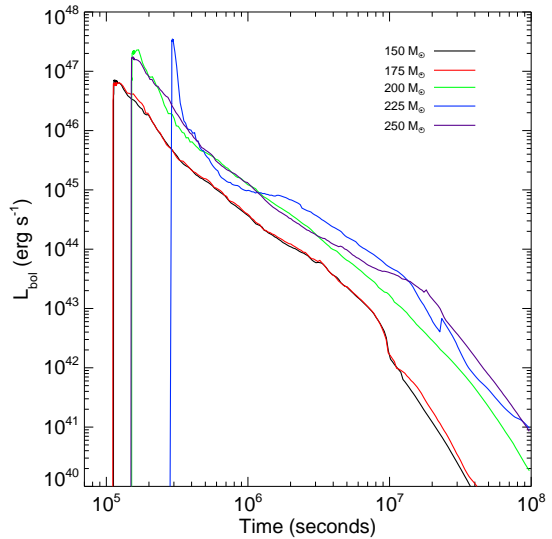}
\caption{Source frame bolometric luminosities for the 5 PI SNe.  The general trend of higher 
luminosity with progenitor mass is due to the larger explosion energies and mass of \Ni\ that 
is synthesized.
\vspace{0.1in}}
\label{fig:lcs}
\end{figure}

Bolometric source frame luminosities are shown for all five explosions in Figure \ref{fig:lcs}. 
The shock is not visible to an external observer before breaking through the surface of the 
star because electrons in the intervening layers scatter the photons.  When it reaches the 
surface, the shock abruptly accelerates in the steep density gradient, becoming even hotter 
and releasing a brilliant pulse of photons with peak luminosities above 10$^{47}$ erg s$^{-1
}$, or 2500 times the luminosity of the Milky Way galaxy.  The transient is mostly X-rays and 
hard UV, and its duration is related to the radius of the star, since photons emitted 
simultaneously from its poles and equator reach an observer at times that are separated by 
the time it takes light to cross the star.  The temperature of the shock at breakout is $\sim$ 
10$^6$ K.  After the initial transient subsides, the luminosities gradually decline from 10$^{
46}$ erg s$^{-1}$ to 10$^{42}$ erg s$^{-1}$ over 3 yr as gamma rays from the decay of \Ni\ 
formed during the explosion downscatter in energy and diffuse out through the ejecta.

Pair-instability SN light curves are powered at early times by the conversion of kinetic energy 
into thermal energy by the shock, and are thus far more luminous than Type Ia and II SNe 
because they have much higher explosion energies.  At later times they are primarily powered 
by radioactive decay, and they remain much brighter than other SNe because they have 
synthesized more \Ni:  up to 50 \Ms\ compared to 0.4--0.8 \Ms\ and $<$ 0.3 \Ms\ in Type Ia 
and II SNe, respectively. They are luminous for longer times (3 yr instead of 3--6 mo for Type 
Ia and II SNe) because radiation diffusion timescales in their more massive ejecta are longer: 
\vspace{0.05in}
\begin{equation}
t_d \sim \kappa^{\frac{1}{2}} {M_{\rm ej}}^{\frac{3}{4}} E^{-\frac{3}{4}}.  \vspace{0.05in}
\end{equation}
Here, $\kappa$ is the average opacity of the ejecta, M$_{\rm ej}$ is the mass of the ejecta, 
and $E$ is the explosion energy.  All five light curves slightly rebrighten at 4 months when 
the photosphere (the surface from which photons stream freely into space) recedes into the 
hot \Ni\ layer deep in the ejecta. This emission peaks at $\sim$ 3000 \AA, and its luminosity 
increases with the mass of \Ni, and hence the mass of the progenitor.  Rebrightening occurs 
later in more massive stars because more time is required for the photosphere to sink into the 
\Ni\ in the frame of the shock.

\begin{figure}
\plotone{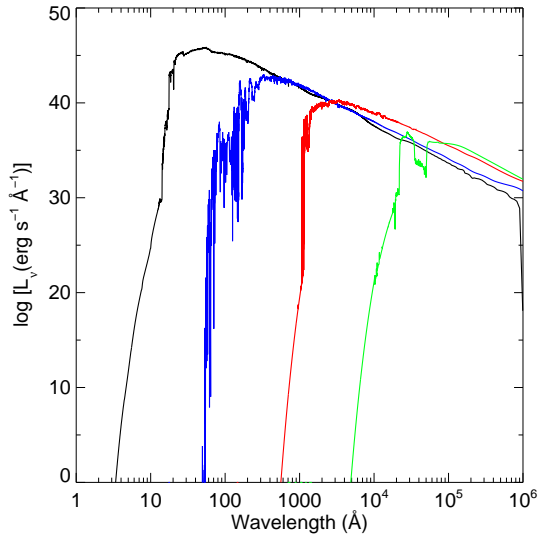}
\caption{Spectral evolution of the 225 M$_{\odot}$ pair instability supernova.  The fireball 
spectra are for 2.91 $\times$ 10$^5$ s (3.4 days, black), 6.13 $\times$ 10$^5$ s (7 days, 
blue), 9.74 $\times$ 10$^6$ s (113 days, red) and 9.70 $\times$ 10$^7$ s (1123 days,
green).  
}
\label{fig:spectra}
\end{figure}

Figure \ref{fig:spectra} shows spectra for the 225 \Ms\ explosion at four times ranging from 
shock breakout to 3 yr.  As the fireball expands it cools, and its spectral peak advances to 
longer wavelengths.  The breakout transient ionizes the envelope shrouding the star, as seen 
at 2.9 $\times$ 10$^5$ s (3.4 days) when the spectrum has far fewer absorption lines than at 
later times when the shock cools and can no longer fully ionize the wind.  As the envelope 
recombines it imprints many more absorption lines on the spectrum, especially at shorter 
wavelengths where line blanketing effectively shears off the spectrum.  Consequently, the 
common practice of fitting blackbodies to bolometric light curves to approximate shock 
spectra can seriously overestimate the NIR flux in the observer frame, because such profiles 
have a significant amount of flux at short wavelengths that is actually removed by the envelope 
and ejecta.  On the other hand, our bolometric luminosities are an order of magnitude higher 
at shock breakout than previous PI SN light curves because the shock crashes into the wind 
and heats it, and because radiation and matter are not forced to be in thermal equilibrium in 
our calculations.  By 3 yr most of the photons in the spectrum are below the ionization limit of 
hydrogen and pass through the envelope, although resonant scattering continues to imprint 
absorption lines on flux below this limit.

\section{JWST and WFIRST Detection Limits}

\begin{figure}
\plotone{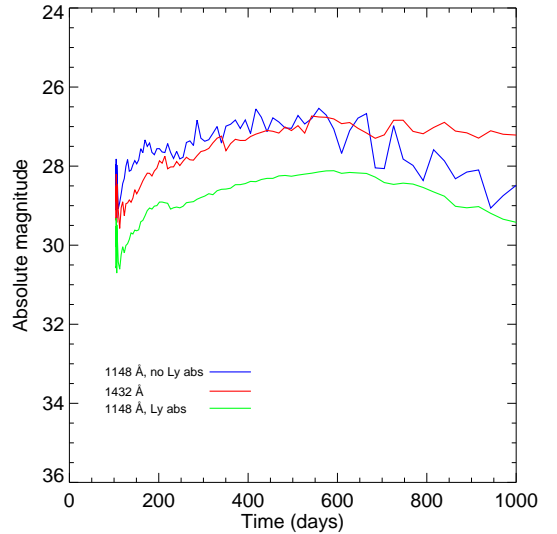}
\caption{NIR light curves with and without Lyman absorption at $z =$ 30.  Blue:  3.56 $\mu$m
(1148 \AA\ in the source frame), no absorption.  Green:  3.56 $\mu$m with absorption.  Red:  
4.44 $\mu$m (1432 \AA\ in the source frame).
}
\label{fig:Lya}
\end{figure}

At high redshift, Lyman absorption will absorb most luminosity at wavelengths far blueward 
of 1216 \AA.  However, it could be possible that the fireball is brighter a little blueward of the 
Lyman limit rather than a little redward, even with absorption.  To determine in which 
\textit{JWST} NIRCam filter an explosion is brightest at a given redshift, we calculate its NIR 
signal at wavelengths above and below the Lyman limit in the observer frame.  At $z =$ 30 
we find that although all five SNe are brighter just below 1216 \AA\ in the source frame, IGM 
absorption reduces these luminosities to magnitudes below those just redward of the limit for
which there is no absorption, as we show for the 225 \Ms\ PI SN in Figure \ref{fig:Lya}.  The 
explosion is brighter at 1148 \AA\ in the source frame (3.56 $\mu$m in the observer frame) at 
early times than at 1432 \AA\ (4.44 $\mu$m in the observer frame), but after IGM absorption 
is taken into account the event is brighter at 4.44 $\mu$m in the observer frame.

\begin{figure*}
\plottwo{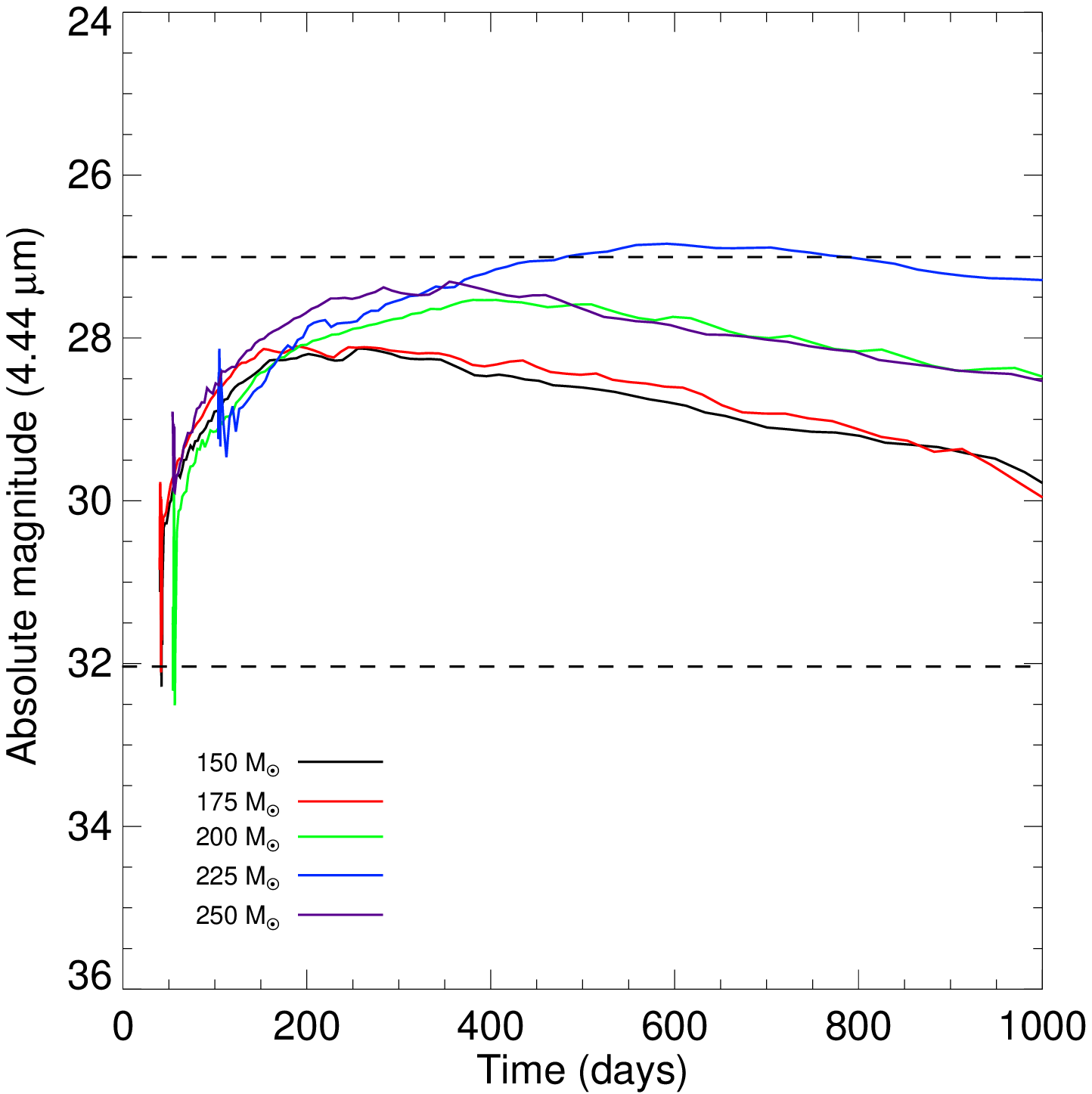}{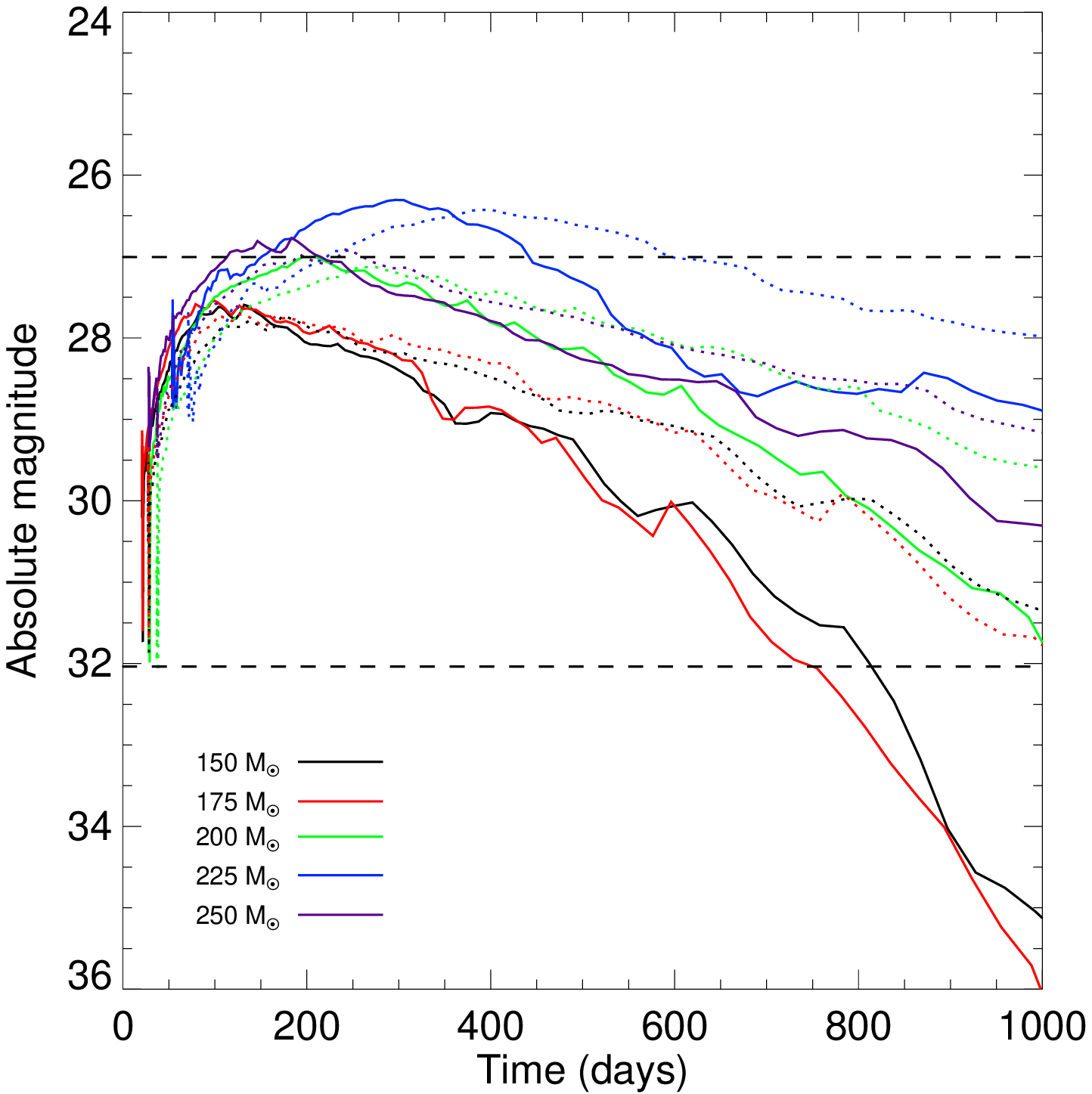}
\caption{NIR light curves for five Pop III PI SNe.  Left:  NIR magnitudes at 4.44 $\mu$m at $z =$ 
30, 100 Myr after the big bang. The horizontal dashed lines at mag 32 and 27 are \textit{JWST} 
and \textit{WFIRST} detection limits, respectively.  The explosions remain above the NIRCam 
detection limit for over 1,000 days.  Right:  NIR light curves at 2.0 $\mu$m at $z = 15$ (solid)  
and $z =$ 20 (dotted). \vspace{0.1in}}
\label{fig:NIR}
\end{figure*}

In the left panel of Figure \ref{fig:NIR} we show NIR light curves for all five Pop III SNe at 4.44 
$\mu$m in \textit{JWST} at $z =$ 30, or 100 Myr after the big bang.  The detection limit for this 
filter is AB magnitude 32, so all five explosions will be visible for over a thousand days.  Their 
light curves do not cross above the detection threshold until 50 days (or 1.5 days in the source 
frame at this redshift), so the most luminous component of the explosion, shock breakout, 
cannot be seen from Earth. The X-rays and extreme UV in the pulse are all redshifted into the 
UV and absorbed by neutral hydrogen.  

The NIR flux in the observer frame exhibits significantly more variability than the bolometric flux 
because of the expansion and cooling of the fireball, and this is essential to identifying these 
events as explosive transients.  The flux rises much more quickly than it falls, so it is easiest to 
detect Pop III PI SNe in their initial stages.  Nonetheless, they exhibit sufficient variability over 
likely protogalactic survey times of 1--5 yr to be identified at later times as well.  This variation 
over timescales of $\sim$1000 days will also allow them to be discriminated from primeval 
galaxies, in spite of their probable overlap in color-color space.  Although recent calculations 
show that roughly one Pop III PI SN will be present in any given \textit{JWST} deep field \citep{
tss09,hum12,jdk12}, far greater numbers could be found in proposed all-sky NIR surveys with 
\textit{WFIRST}.  With a sensitivity of AB magnitude 27 at 2.2 $\mu$m, \textit{WFIRST} will see 
such explosions at $z =$ 15--20, as we show in the right panel of Figure \ref{fig:NIR}.  Such 
redshifts may be optimal for detecting PI SNe because of the rise of UV background fields from 
the first generations of Pop III stars.  This background is thought to delay primordial star 
formation in halos until they have grown to larger masses, forming more massive stars at slightly 
lower redshifts \citep{wa07,on08}.  Star formation rates extracted from large-scale cosmological 
simulations of early structure formation that include UV backgrounds \citep{jdk12} suggest that 
WFIRST will detect up to 10$^3$ PI SNe per year at $15< z < 20$.  Detections of this volume will 
enable the first major surveys of the Pop III initial mass function (IMF) \citep[see][for discussions
on strategies for detecting superluminous Type IIn SNe in all-sky surveys]{tet12,moriya12}.

Our calculations definitively establish that \textit{JWST} and \textit{WFIRST} will be able to detect 
Pop III supernovae at the era of first light.  In addition to unveiling the nature of primordial stars 
and constraining scenarios for early cosmological reionization and chemical enrichment, 
detections of Pop III supernovae at slightly lower redshifts ($z \sim$ 10--15) will probe the era of 
primitive galaxy formation, revealing their positions on the sky when they might otherwise have 
eluded detection by \textit{JWST}.  The discovery of the first cosmic explosions will be one of the 
most spectacular results in extragalactic astronomy in the coming decade, and open the first 
window on the end of the cosmic Dark Ages.

\acknowledgments

The authors thank the anonymous referee, whose comments improved the quality of this paper.
DJW was supported by the Bruce and Astrid McWilliams Center for Cosmology at Carnegie 
Mellon University. DEH acknowledges support from the National Science Foundation CAREER 
grant PHY-1151836.  AH was supported by the US Department of Energy under contracts 
DE-FC02-01ER41176, FC02-09ER41618 (SciDAC), and DE-FG02-87ER40328.  SEW was 
supported by the National Science Foundation grant AST 0909129 and the NASA Theory 
Program grant NNX09AK36G.  MS thanks Marcia Rieke for making available the NIRCam filter 
curves, and was partially supported by NASA JWST grant NAG5-12458.  Work at LANL was 
done under the auspices of the National Nuclear Security Administration of the U.S. Department 
of Energy at Los Alamos National Laboratory under Contract No. DE-AC52-06NA25396.  

%\bibliographystyle{apj}
%\bibliography{refs}

\end{document}